\documentclass[prd,aps,preprint,showpacs]{revtex4}
\usepackage{epsfig}
\usepackage{axodraw}

\begin{document}

\begin{titlepage}

\voffset 1.5cm
\preprint{KIAS-P03023, hep-ph/0304069}

\title{
\large
Testing Higgs Triplet Model and Neutrino Mass Patterns
}

\author{
 \vspace{2ex}
  Eung Jin Chun, Kang Young Lee  and Seong Chan Park}

\affiliation{
\vspace{1ex}
 Korea Institute for Advanced Study, 207-43 Cheongryangri2-dong,
 Dongdaemun-gu, Seoul 130-722, Korea
\vspace{2cm}  }


\begin{abstract}
The observed neutrino oscillation data might be explained  by
new physics at a TeV scale, which is testable in the future
experiments.  Among various possibilities,  the low-energy Higgs 
triplet model is a prime candidate of such new physics since it
predicts clean signatures of lepton flavor violating processes 
directly related to  the neutrino masses and mixing.
It is discussed how various neutrino mass patterns can be 
discriminated by examining the lepton flavor violating decays 
of charged leptons as well as the collider signatures of 
a doubly  charged Higgs boson in the model.
\end{abstract}


\maketitle

\end{titlepage}


\voffset 0cm

\section{Introduction}

The atmospheric, solar and reactor neutrino experiments
\cite{SK,SNO,CHZ,KamL} have firmly established the picture of three active
neutrino oscillations, and provided  us important information on two
neutrino mass-squared differences and three mixing angles.
Taking the most favorable parameter region of the solar neutrino
oscillation (so-called LMA I), we have
\begin{eqnarray}
&&\Delta m^2_{atm}=(1.1-4.8)\times10^{-3}\, \mbox{eV}^2, \quad
\sin^2\theta_{atm} = 0.3-0.7 \,,
\nonumber\\
&&\Delta m^2_{sol}=(0.5-1.0)\times10^{-4}\, \mbox{eV}^2,  \quad~
\sin^2\theta_{sol} = 0.24-0.44 \,,
\end{eqnarray}
and the limit of $\sin^2\theta_{chooz} < 0.038$ coming from
the non-observation of $\nu_\mu\to \nu_e$ oscillation
in the CHOOZ and atmospheric neutrino data \cite{CHZ,SK}.

Given such new experimental inputs,
we could hope for uncovering new physics beyond the
standard model, which must explain the observed neutrino data.
In this regard, a ``low-energy'' model for neutrino masses and mixing
is of particular interest since it may be tested in the future
experiments observing lepton flavor violating processes in accelerators.
A typical example of such a model would be the supersymmetric standard
model with {\it R-parity violation} in which the flavor structure of
neutrino mass matrix could be probed through the decay of the 
lightest supersymmetric particle \cite{rpv}.
Another example is the Zee model and its variations \cite{babu1}
which rely on {\it radiative mechanism} of neutrino mass generation.

In this paper, we consider the {\it Higgs triplet model} in which a
triplet scalar field $\Delta=(\Delta^{++},\Delta^+,\Delta^0)$
with the mass $M$ is introduced
to have the following renormalizable couplings;
\begin{equation}
{\cal L}_\Delta = {1\over\sqrt{2}}[ f_{ij} L_i L_j \Delta + 
 \mu\Phi\Phi \Delta + h.c.] -M^2 |\Delta|^2\,,
\end{equation}
where $L_i=(\nu_i,l_i)_L$ is the left-handed lepton doublet and
$\Phi=(\phi^0,\phi^-)$
is the standard model Higgs doublet.  Due to the ``$\mu$'' term in the
above equation, the neutral component $\Delta^0$ of the triplet
gets the vacuum expectation value (VEV), $v_\Delta =
\mu v_\Phi^2/2M^2$ where $v_\Phi= \langle \Phi^0 \rangle = 246$ GeV.
This leads to the neutrino mass matrix,
\begin{equation}
M^\nu_{ij}=  f_{ij}  v_\Delta \,.
\end{equation}
We are interested in the possibility of the light triplet Higgs bosons, 
namely $M \sim v_\Phi$, so that observations of various lepton flavor
violating processes can provide a probe for the neutrino
masses and mixing through the relation (3), and thus  a direct test
of the model. In this ``low-energy triplet Higgs model'',
the small parameters $f$ and $\xi \equiv v_\Delta/v_\Phi$ are required;
\begin{equation}
f_{ij} \xi  \sim 10^{-12}
\end{equation}
for $M^\nu_{ij}\sim 0.3$ eV.
We will see later that such a smallness could be understood by
a radiative mechanism.  Here, let us note that 
we are interested in the case of very small $\xi$, say 
$\xi \lesssim 10^{-6}$, so that the condition of 
$\rho=m_Z^2/m_W^2 c_W^2\simeq 1$ is simply satisfied in our consideration.

Phenomenological consequences of  low-energy triplet Higgs bosons
have been studied extensively in the past, in particular, 
centering around the exotic signatures of a doubly charged Higgs boson,
$\Delta^{\pm\pm}$ \cite{gunion,xyz,rizzo,lhc,ma1,olds}.
The main purpose of this work is to investigate how the observation
of such phenomena can test the pattern of the neutrino masses
and mixing.  For this to happen, we will mostly assume that
$f \gtrsim \xi$ to detect the lepton flavor violating processes
induced by the coupling $f$.  
This paper is organized as follows.  In section 2, we derive the
flavor structure of the bileptonic couplings $f_{ij}$ depending on
the acceptable neutrino mass patterns, based on which the
observability of rare lepton decays such as $\mu \to e \gamma$,
$\mu\to 3e$ and $\tau\to 3l$ will be discussed.
In section 3, we will consider the production and decays of doubly 
charged Higgs bosons in colliders from which some information on 
the couplings $f$ can be obtained.  We will see when the
collider effects of the coupling $f$ can be observed in relation
to the above discussion.  Then, we examine how the neutrino mass 
patterns can be discriminated through the observation of 
$\Delta^{\pm\pm}$ decays.
In section 4, we present a model in which the smallness of the
couplings $f$ and $\mu$ is explained by a radiative generation at
two-loop level.  We conclude in section 5.

\section{Neutrino mass patterns and low-energy
lepton flavor violation}

Current neutrino data (1) give us the following neutrino mixing matrix;
\begin{equation}
U \approx \pmatrix{ c_3 & s_3 & s_2 \cr
   -{s_3\over\sqrt{2}} & {c_3\over\sqrt{2}} & {1\over\sqrt{2}} \cr
  {s_3\over\sqrt{2}} & -{c_3\over\sqrt{2}} & {1\over\sqrt{2}} \cr}
\end{equation}
in the leading approximation where we put $c_2\simeq1$, $c_1\simeq
s_1\simeq 1/\sqrt{2}$.   Note that the mixing angles in Eq.~(1) can
be identified as $\theta_{atm} \approx \theta_1$,
$\theta_{sol} \approx \theta_3$ and
$\theta_{chooz} \approx \theta_2$.  Then, the flavor structure of
the coupling $f$ can be determined simply by $f
\propto M^\nu \approx U \mbox{ diag}(m_1,m_2,m_3) U^T$.
In the below, we will show the ratios;
\begin{eqnarray}
[ff^\dagger] &\equiv&
(ff^\dagger)_{11}: (ff^\dagger)_{22}: (ff^\dagger)_{33} :
   (ff^\dagger)_{12}: (ff^\dagger)_{13}: (ff^\dagger)_{23}\,,
   \nonumber\\
\mbox{and}\qquad
 [f]&\equiv& f_{11}: f_{22}: f_{33} : f_{12}: f_{13}: f_{23} \,.
 \nonumber
\end{eqnarray}
Given the information on $\Delta m^2$ (1), one has a variety of 
possibilities for the neutrino mass eigenvalues.  
Assuming CP conservation, the following different patterns can be allowed:

\noindent
 (i) Hierarchy with $m_1<m_2<m_3$ which gives
\begin{eqnarray} \label{HI}
{}[ff^\dagger] &=&
   ~~(s_2^2 + r s_3^2) : {1\over2} : {1\over2} :
   {1\over\sqrt{2}}(s_2+{r\over2}\sin2\theta_3):
   {1\over\sqrt{2}}(s_2-{r\over2}\sin2\theta_3): {1\over2} \\
\mbox{HI}~~ [f] ~~ &= &
 (s_2^2 + \sqrt{r} s_3^2) : {1\over2} : {1\over2} :
   {1\over\sqrt{2}}(s_2+{\sqrt{r}\over2}\sin2\theta_3):
   {1\over\sqrt{2}}(s_2-{\sqrt{r}\over2}\sin2\theta_3): {1 \over2}
\end{eqnarray}
where $r\equiv \Delta m^2_{atm}/\Delta m^2_{sol}$ which is in the
range of $[0.01-0.1]$ as in Eq.~(1).

\noindent
 (ii) Inverse Hierarchy with $m_1 \simeq m_2 \gg
m_3$ (IN1) and $m_1=-m_2\gg m_3$ (IN2) resulting in
\begin{eqnarray} \label{IN}
{}[ff^\dagger] &=&
   1 : {1\over2} : {1\over2} :
   {1\over\sqrt{2}}(s_2+{r\over2}\sin2\theta_3):
   {1\over\sqrt{2}}(s_2-{r\over2}\sin2\theta_3): {1\over2} \\
\mbox{IN1} ~~[f] ~~ &= &
   1 : {1\over2} : {1\over2} :
   {1\over\sqrt{2}}(s_2-{r\over4}\sin2\theta_3):
   {1\over\sqrt{2}}(s_2+{r\over4}\sin2\theta_3): {1\over2}  \\
\mbox{IN2} ~~ [f] ~~ &= &
   \cos2\theta_3 : {1\over2}(\cos2\theta_3-s_2 \sin2\theta_3) :
    {1\over2}(\cos2\theta_3+s_2 \sin2\theta_3) :  \nonumber\\
&&   {1\over\sqrt{2}}\sin2\theta_3:
   {1\over\sqrt{2}}\sin2\theta_3: {1\over2}\cos2\theta_3
\end{eqnarray}

\noindent
 (iii) Degeneracy with $m_1 \simeq m_2 \simeq
m_3$ (DG1), $m_1 \simeq m_2 \simeq -m_3$ (DG2), $m_1 \simeq -m_2
\simeq m_3$ (DG3), $m_1 \simeq -m_2 \simeq -m_3$ (DG4) yielding
\begin{eqnarray} \label{DG}
{}[ff^\dagger] &=&
   1 : 1: 1 :
   {R\over\sqrt{2}}(s_2+{r\over2}\sin2\theta_3):
   {R\over\sqrt{2}}(s_2-{r\over2}\sin2\theta_3): {R\over2} \\
\mbox{DG1} ~~[f] ~~ &= &
   1 : 1 : 1 :
   {R\over2\sqrt{2}}(s_2+{r\over2}\sin2\theta_3):
   {R\over2\sqrt{2}}(s_2-{r\over2}\sin2\theta_3): {R\over4}  \\
\mbox{DG2} ~~[f] ~~ &= &
   1: s_2^2+\cos2\theta_1 -{R\over4}:
   s_2^2-\cos2\theta_1 -{R\over4}: \nonumber\\
&&    \sqrt{2}(s_2-{r\over4} \sin2\theta_3) :
    \sqrt{2}(s_2+{r\over4} \sin2\theta_3) :  1  \\
\mbox{DG3}~~[f] ~~ &= &
   \cos2\theta_3: s_3^2 + s_2 \sin2\theta_3:
   s_3^2 - s_2 \sin2\theta_3:  \nonumber\\
&&   {1\over\sqrt{2}}(\sin2\theta_3 - 2s_2 s_3^2):
   {1\over\sqrt{2}}(\sin2\theta_3 + 2s_2 s_3^2): c_3^2 \\
\mbox{DG4}~~[f] ~~ &= &
   \cos2\theta_3: c_3^2 - s_2 \sin2\theta_3:
  c_3^2 + s_2 \sin2\theta_3:  \nonumber\\
&&   {1\over\sqrt{2}}(\sin2\theta_3 + 2s_2 c_3^2):
   {1\over\sqrt{2}}(\sin2\theta_3 - 2s_2 c_3^2): s_3^2
\end{eqnarray}
where $R\equiv \Delta m^2_{atm}/m_1^2$.  Since the recent  WMAP 
results put a limit of $m_1< 0.23$ eV \cite{wmap},
the ratio $R$ has to be larger than about $0.02$.

The schematic form of the bilepton couplings (2) can be written explicitly
as
\begin{eqnarray}
 {\cal L} &=&
 {1\over\sqrt{2}} f_{ij}\, \bar{L}^c_i\, i\tau_2 \mathbf{\Delta}\, L_j + h.c.
 \\
 &=&  -{1\over2} f_{ij} \left[ \sqrt{2}\, \bar{l}^c_i P_L l_j \Delta^{++}
    + (\bar{l}^c_i P_L \nu_j + \bar{l}^c_j P_L \nu_i) \Delta^{+}
    -\sqrt{2}\,  \bar{\nu}^c_i P_L  \nu_j \Delta^{0} + h.c. \right] \,,
    \nonumber
\end{eqnarray}
where  we used the matrix form of the triplet field;
$$\mathbf{\Delta}= \pmatrix{ {\Delta^+\over\sqrt{2}} & \Delta^{++} \cr
                             \Delta^0 & -{\Delta^+\over\sqrt{2}} \cr} .
$$
The above Lagrangian induces the tri-leptonic and radiative decays
of a charged lepton at tree and one-loop level, respectively \cite{olds}.
Let us now discuss the observational possibilities of such 
lepton flavor violating decays of muon or tau in the triplet Higgs model.  
Table I shows the
current limits on the products of couplings for various decay modes,
and their future experimental sensitivities.
For the discovery of some lepton flavor violating decay modes, one needs
\begin{eqnarray}
f_{11}f_{12}~ &>& 3.0\times10^{-8} x_\Delta
\quad \mbox{for}\quad \mu \to 3e,
\nonumber\\
(ff^\dagger)_{12} &>& 3.5\times10^{-6} x_\Delta
\quad\mbox{for}\quad \mu \to e \gamma,
\\
f_{ij} f_{k3}~ &\gtrsim& 2.3\times10^{-4} x_\Delta
\quad\mbox{for} \quad \tau \to 3l.
\nonumber
\end{eqnarray}
where $i,j,k=1,2$ as indicated in Table I.

In the cases of (IN2), (DG3) and (DG4), neither $\mu\to e\gamma$
nor $\tau \to 3l$ can be observed as the strong constraint from the
$\mu \to 3e$ pushes them outside the future experimental sensitivity.
To see this, let us note that
$f_{11}f_{12} \propto \sin2\theta_3\cos2\theta_3/\sqrt{2}$ from
Eqs.~(10), (14) and (15), and $\cos2\theta_3>0.1$ from Eq.~(1),
which shows that
\begin{eqnarray} \label{no1}
 f_{ij}f_{k3} &<& {f_{11}f_{12}\over \cos2\theta_3} < 10^{-5} x_\Delta
 \nonumber\\
 (ff^\dagger)_{12} &=&
  { (R) s_2 \over \cos2\theta_3 \sin2\theta_3} f_{11}f_{12}
  < 2\times 10^{-6} x_\Delta 
\end{eqnarray}
where $R$ has to be included in the (DG) case.
The situation can be different in other cases
where one has the following relations for the ratio  
$f_{11}f_{12}: (ff^\dagger)_{12} : f_{ii} f_{23}$;
\begin{eqnarray}
\mbox{(HI)}& & {2\sqrt{2}}(s_2+{\sqrt{r}\over2}\sin2\theta_3)\sqrt{r}s_3^2 :
              {2\sqrt{2}}(s_2+{r\over2}\sin2\theta_3)  : 1 \nonumber\\
\mbox{(IN1)}& & \sqrt{2}(s_2-{r\over4}\sin2\theta_3):
              \sqrt{2}(s_2+{r\over2}\sin2\theta_3)  : 1 \nonumber\\
\mbox{(DG1)}& & \sqrt{2}(s_2+{r\over2}\sin2\theta_3):
              {2\sqrt{2}}(s_2+{r\over2}\sin2\theta_3)  : 1 \nonumber\\
\mbox{(DG2)}& & \sqrt{2}(s_2-{r\over4}\sin2\theta_3):
              {R\over\sqrt{2}}(s_2+{r\over2}\sin2\theta_3)  : 1
\end{eqnarray}
where $f_{ii}=f_{22}$ for (HI) and $f_{11}$ otherwise.
\begin{table}
\begin{tabular}{|c|c|c|c|}
\hline Mode & ~Current limit \cite{pdg,yusa}~
     & ~Future sensitivity \cite{fut1,yusa}~
     &~ Bound on the couplings~ \\
\hline $\mu\to e \gamma$ & $1.2\times10^{-11}$ & $\sim10^{-14}$
     & $(f f^\dagger)_{12} < 1.2\times10^{-4}\, x_\Delta$ \\
$\tau\to e \gamma$ & $2.7\times10^{-6}$ & $\sim10^{-8}$
     & $(ff^\dagger)_{13} < 1.3\times10^{-1}\, x_\Delta$ \\
$\tau\to \mu \gamma$ & $0.6\times10^{-6}$ & $\sim10^{-8}$
     & $(ff^\dagger)_{23} < 6.1\times10^{-2}\, x_\Delta$ \\
$\mu\to \bar{e}ee$ & $1.0\times10^{-12}$ & $\sim10^{-15}$
     & $f_{11}f_{12} < 9.3\times10^{-7}\, x_\Delta$ \\
$\tau\to \bar{e}ee$ & $2.7\times10^{-7}$ & $\sim10^{-8}$
     & $f_{11}f_{13} < 1.1\times10^{-3}\, x_\Delta$ \\
$\tau\to \bar{e}e\mu$ & $2.4\times10^{-7}$ & $\sim10^{-8}$
     & $f_{12}f_{13} < 1.5\times10^{-3}\, x_\Delta$ \\
$\tau\to \bar{e}\mu\mu$ & $3.2\times10^{-7}$ & $\sim10^{-8}$
     & $f_{22}f_{13} < 1.2\times10^{-3}\, x_\Delta$ \\
$\tau\to \bar{\mu}ee$ & $2.8\times10^{-7}$ & $\sim10^{-8}$
     & $f_{11}f_{23} < 1.2\times10^{-3}\, x_\Delta$ \\
$\tau\to \bar{\mu}e\mu$ & $3.1\times10^{-7}$ & $\sim10^{-8}$
     & $f_{12}f_{23} < 1.7\times10^{-3}\, x_\Delta$ \\
~$\tau\to \bar{\mu}\mu\mu$~ & $3.8\times10^{-7}$ & $\sim10^{-8}$
     & $f_{22}f_{23} < 1.4\times10^{-3}\, x_\Delta$ \\
\hline
\end{tabular}
\caption{The experimental limits on the branching ratios of
various modes and the corresponding upper bounds on the product of
couplings taking $x_\Delta = (M_{\Delta}/200 {\rm GeV})^2$.}
\end{table}
From this, one can see that the decay modes other than $\mu \to 3e$ can
be seen only if the coupling $f_{12}$ is made small and thus the following
relation is fulfilled;
$s_2 \approx -\sqrt{r} \sin2\theta_3/2$ (HI),
$s_2\approx r \sin2\theta_3/4$ (IN1),
$s_2\approx - r \sin2\theta_3/2$ (DG1) or
$s_2\approx  r \sin2\theta_3/8$ (DG2).
In this case,  one predicts
\begin{itemize}
\item (HI)~~ $B(\tau \to \bar{\mu}\mu\mu) : B(\mu\to e\gamma)
 = 1 : 8.6\times10^{-3} r\sin^22\theta_3$
\item (IN1)~ $B(\tau \to \bar{\mu}ee) : B(\tau\to\bar{\mu}\mu\mu)
 : B(\mu \to e\gamma)
 = 1 : 0.5 : 4.8\times10^{-3} r^2 \sin^22\theta_3$
\item (DG1) $B(\tau \to \bar{\mu}ee):  B(\tau \to \bar{\mu}\mu\mu)
 = 1 : 1 $
\item (DG2)  $B(\tau \to \bar{\mu}ee) :  B(\mu \to e\gamma)
 = 1 : 4.8\times10^{-3} R^2 r^2 \sin^22\theta_3 $
\end{itemize}
An ideal case is to observe both $\tau\to 3l$ and $\mu\to e\gamma$
decays which will enable us to discriminate the different
mass patterns.

\section{collider test: production and decays of Higgs triplet}

Some of striking collider signals in the triplet Higgs model comes
from the decays of a doubly charged Higgs boson,
such as $\Delta^{--} \to l_i l_j, W^-W^-$,
which have been studied extensively in the past years
\cite{gunion,xyz,rizzo,lhc,ma1,olds}.
We are interested in the situation that
the decays $\Delta^{--} \to l_i l_j$ are sizable so that the neutrino
mass structure can be tested in colliders.   Depending on the masses
of the triplet components, the fast decay process like
$\Delta^{--} \to \Delta^- W^{(*)-}$ through gauge interactions can happen to
over-dominate any other processes of our interest.
The mass splitting among the triplet components arises upon the electroweak
symmetry breaking and thus is of the order $M_W$.
In order to study the mass spectrum and decay processes
of the triplet Higgs bosons,
let us first consider the most general scalar potential
for a doublet and a triplet Higgs boson:
\newcommand{\bfD}{\mathbf{\Delta}}
\begin{eqnarray}
V&=& m^2 (\Phi^\dagger\Phi) + \lambda_1 (\Phi^\dagger\Phi)^2
  + M^2 \mbox{Tr}(\bfD^\dagger\bfD)
  + \lambda_2 [\mbox{Tr}(\bfD^\dagger\bfD)]^2
  + \lambda_3 \mbox{Det}(\bfD^\dagger\bfD) \nonumber\\
&&  + \lambda_4 (\Phi^\dagger\Phi)\mbox{Tr}(\bfD^\dagger\bfD)
  + \lambda_5 (\Phi^\dagger \tau_i\Phi)\mbox{Tr}(\bfD^\dagger \tau_i \bfD)
  + {1\over\sqrt{2}}\mu  (\Phi^Ti\tau_2 \bfD \Phi)  +h.c.
\end{eqnarray}
Note that the triplet VEV is given by
$v_\Delta= \mu v_\Phi^2/2M_{\Delta^0}^2$.  In this theory,
the mass eigenstates consist of $\Delta^{++}$, $H^+$,
$H^0$,  $A^0$ and $h^0$.  Under the condition that $|\xi| \ll 1$,
the first five states are mainly from the triplet sector and
the last from the doublet sector.  The approximate mass diagonalizations
are given as follows. For
the neutral pseudoscalar and charged scalar parts,
\begin{eqnarray}
 \phi^0_I = G^0 - 2 \xi A^0 \;, \qquad
 \phi^+ = G^+ + \sqrt{2} \xi H^+  \nonumber\\
 \Delta^0_I= A^0 + 2 \xi G^0 \;, \qquad
 \Delta^+= H^+ - \sqrt{2} \xi G^+
\end{eqnarray}
where $G^0$ and $G^+$ are the Goldstone modes, and
for the neutral scalar part,
\begin{eqnarray}
 \phi^0_R &=& h^0 - a \xi \, H^0 \,, \nonumber\\
 \Delta^0_R&=& H^0 + a \xi \, h^0
\end{eqnarray}
where $ a = 2 + 4 (4\lambda_1-\lambda_4-\lambda_5) m^2_{W}/
   g^2(m^2_{H^0}-m^2_{h^0}) $.
The masses of the Higgs bosons are
\begin{eqnarray} \label{massD}
 M^2_{\Delta^{\pm\pm}} &=& M^2 + 2{\lambda_4 -\lambda_5 \over g^2 } M^2_{W}
 \nonumber\\
 M^2_{H^{\pm}} &=& M_{\Delta^{\pm\pm}}^2  + 2 {\lambda_5 \over g^2} M^2_{W}
 \\
 M^2_{H^0, A^0} &=&  M^2_{H^\pm} +
    2{\lambda_5 \over g^2} M^2_{W} \,.
 \nonumber
\end{eqnarray}
The mass of $h^0$ is
given by $m_{h^0}^2=4\lambda_1 v_\Phi^2$ as usual.

When $\lambda_5>0$, we have 
$M_{\Delta^{\pm\pm}} < M_{H^\pm} < M_{H^0,A^0}$, so that the doubly charged
Higgs boson $\Delta^{--}$ can only decay to $l_i l_j$ or $W^- W^-$
through the following interactions;
\begin{equation}
 {\cal L} = {1\over\sqrt{2}} \left[ f_{ij}\, \bar{l^c}_i P_L l_j
  +  g\xi M_W\, W^- W^-  \right] \Delta^{++} + h.c.
\end{equation}
The corresponding decay rates are
\begin{eqnarray}
\Gamma(\Delta^{--}\to l_i l_j) &=& S {f_{ij}^2 \over 16 \pi} 
     M_{\Delta^{\pm\pm}} \nonumber \\
\Gamma(\Delta^{--}\to W W) &=& {\alpha_2 \xi^2 \over32}
           {M_{\Delta^{\pm\pm}}^3\over M_W^2}
	      (1-4r_W+12r^2_W) (1-4r_W)^{1/2}
\end{eqnarray}
where $S=2\,(1)$ for $i\neq j\, (i=j)$
and $r_W=M_W^2/M_{\Delta^{\pm\pm}}^2$.  In this case, the heavier states
$H^+$, $H^0$ and $A^0$ can have the decay modes;
$H^0,A^0 \to H^+ W^{(*)-}$ and $H^+\to \Delta^{++} W^{(*)-}$ 
leading to the production of $\Delta^{\pm\pm}$.

When $\lambda_5<0$, one has $M_{\Delta^{\pm\pm}} > M_{H^\pm} >M_{H^0,A^0}$.
In this case,  the decay processes
of $\Delta^{--} \to H^- W^-$ and $H^- \to H^0(A^0)\, W^-$ can be allowed
through the usual gauge interactions;
\newcommand{\lrpartial}
{\overleftarrow{\partial}\hspace{-2.3ex}\overrightarrow{\partial}}
\newcommand{\mE}{E\hspace{-1.3ex}\slash}
\begin{equation}
{\cal L}= i g W^+ [ H^+ \lrpartial \Delta^{--}
     + {1\over\sqrt{2}} H^0 \lrpartial H^-
     + {i\over\sqrt{2}} A^0 \lrpartial H^- ] + h.c. \,,
\end{equation}
giving rise to the decay rate
\begin{equation}
 \Gamma(\Delta^{--} \to H^- W^- ) = {g^2\over 8\pi} M_W
           \left[1+{2y^2-y-1 \over 2} r_W\right]
           \left[{(y+1)^2\over4} r_W-1 \right]^{1/2}
\end{equation}
where $y\equiv 2|\lambda_5|/g^2$.  
This can be rewritten as 
$ \Gamma(\Delta^{--} \to H^- W^- )
= (5 \sqrt{2}g^2/8\pi) M_W \delta^{1/2}$
in the limit of $\delta\equiv (M_{\Delta^{\pm\pm}}-M_{H^\pm}-M_W)/M_W \to 0$
that is, $y+1\to 2 r_W^{-1/2}$.
To suppress the deday mode of Eq.~(27), we will require 
$M_{\Delta^{\pm\pm}}< M_{H^\pm} + M_W$, that is, $M_{\Delta^{\pm\pm}} >
{(y+1)\over2}M_W$.  For $M_{\Delta^{\pm\pm}}= 200$ GeV, it implies
$ |\lambda_5|<0.89$.  Thus, the decay 
$\Delta^{--} \to H^- W^-$ is forbidden unless the coupling $\lambda_5$ is
extremely large.  Now, the off-shell production of $W$, 
$\Delta^{--} \to H^- W^{*-}$, is allowed to have the rate;
\begin{equation}
 \Gamma(\Delta^{--} \to H^{-} W^{*-} ) \approx  {3 G_F^2 \over 40 \pi^3}
  {y^5 M_W^{10}\over M_{\Delta^{\pm\pm}}^5}
\end{equation}
in the leading term of $y M_W^2$.  With the further requirement of  
$\Gamma(\Delta^{--} \to H^{-} W^{*-} ) < \Gamma(\Delta^{--} \to l_i l_j )$,
we limit ourselves in the parameter space satisfying 
\begin{equation}
 |\lambda_5| < 0.16
  \left( M_{\Delta^{\pm\pm}} \over 200 \mbox{ GeV}\right)^{6/5}
        \left( f_{ij} \over 10^{-3} \right)^{2/5} \,.
\end{equation}

Here, let us remark that, after the diagonalization in Eqs.~(21) and (22),
we also get couplings for the interactions,
$H^+ \to u\bar{d}, h^0 W^+, Z W^+$ and $H^0, A^0 \to f\bar{f},
W^+ W^-, ZZ, h^0 h^0, Zh^0$, all proportional to $\xi$,
and thus they should be considered as well if $f\sim \xi$.

Before going to our main discussion, let us note that the triplet
Higgs decay is short enough to occur inside colliders.
Assuming Eq.~(25) as the main decay rates and 
recalling $\sum_{ij} f^2_{ij} \propto \mbox{Tr}(M^2_\nu)$
where $M^\nu_{ij}=f_{ij} \xi v_\Phi$, one obtains the following form 
of the total decay rate:
\begin{equation}
 \Gamma_{\Delta^{\pm\pm}}=  M_{\Delta^{\pm\pm}} \left(
  {1\over 16\pi } {\bar{m}^2 \over  \xi^2 v_\Phi^2}
 + {\alpha_2 \over32} {\xi^2 \over r_W}
	      (1-4r_W+12r^2_W) (1-4r_W)^{1/2}  \right)
\end{equation}
where $\bar{m}^2\equiv \sum_i m_i^2 $.
When  $M_{\Delta^{\pm\pm}} > 2 M_W$,  one finds the minimum value of 
the total decay rate given by 
$$ \Gamma_{\Delta^{\pm\pm}}|_{min}= 
 {1\over 8\pi } {M_{\Delta^{\pm\pm}}\bar{m}^2 \over \hat{\xi}^2 v_\Phi^2}$$
where $\hat{\xi}^2\equiv  (2\sqrt{2}/g) r^{1/2}_W
(\bar{m}/v_\Phi) (1-4r_W+12 r^2_W)^{-1/2}(1-4r_W)^{-1/4}$.
Taking $\bar{m}=0.05$ eV and $M_{\Delta^{\pm\pm}}=200$ GeV, we obtain
$\Gamma_{\pm\pm}|_{min}\approx 6\times10^{-13}$ GeV and $\hat{\xi}
\approx 6\times10^{-7}$, leading to $\tau|_{max}
\approx 0.03$ cm.  
When $M_{\Delta^{\pm\pm}} < 2 M_W$, only the first term in Eq.~(30)
contributes and the total decay rate is then $\Gamma > 8\times10^{-14}$ 
GeV for $ M_{\Delta^{\pm\pm}}=100$ GeV and $\xi < 10^{-6}$.
Thus, as far as $\Delta^{--} \to l_i l_j$ are the main decay modes of the 
doubly charged Higgs boson, its decay signal should be observed 
in colliders.

\bigskip

%
%
%
%

\noindent
$\bullet$ Single production of $\Delta^{\pm\pm}$:
$e^+e^- \to e^\pm l^\pm  \Delta^{\mp\mp}$

In the $e^+ e^-$ colliders, an energetic virtual photon emitted from
$e^\pm$ leads to the enhanced $e^\mp \gamma$ scattering producing
$l_i^\pm \Delta^{\mp\mp}$ when a coupling $f_{1i}$ is sizable.  
Adopting the result of Ref.~\cite{rizzo}
with the $p_T$ cut ($p_T=10$ GeV) and neglecting
the final state lepton masses,
we obtain the following pairs of $M_{\Delta^{\pm\pm}}$ and $f_{1i}^2$:
\begin{equation}
\begin{tabular}{c|ccccccc}
\hline
$M_{\Delta^{\pm\pm}}$ 
 (\mbox{GeV}) &~100~&~400~&~600~&~700~&~800~&~850~&~900~\cr
\hline
$f^2_{1i}$ ($10^{-6} x_\Delta$) & 2.8 & 3.4 & 5.4 & 7.6 & 12 &  17  & 29  \cr
\hline
\end{tabular}
\end{equation}
to get the cross-section of $\sigma =0.01$ fb at $\sqrt{s}=1$ TeV.
This corresponds to $N=10$ events for the integrated luminosity $L=1000$/fb.
The cross-section of course scales with $f^2_{1i}$ given the mass
$M_{\Delta^{\pm\pm}}$. 

Let us first consider the cases of (IN2), (DG3) and (DG4) where
the couplings $f_{1i}^2$ are strongly constrained as seen in Eq.~(17).
In each case, we get
\begin{equation}
 (f_{11}^2, f_{12}^2, f_{13}^2) \approx
  (\cot2\theta_3,\, {1\over2}\tan2\theta_3, \,
      {1\over2}\tan2\theta_3)\, \sqrt{2} f_{11}f_{12}
\end{equation}
neglecting a small deviation due to the contribution of $s_2$.
Thus, if $\mu \to 3e$ decay is found near the current experimental limit
and $\theta_3$ is close to 45$^o$,
the final states  $\mu^\pm \Delta^{\mp\mp}$ and
$\tau^\pm \Delta^{\mp\mp}$ could be observed with
$$ N(\mu \Delta) = N(\tau \Delta) $$
for smaller values of the triplet mass, say $M_{\Delta^{\pm\pm}}<700$ GeV.

In the cases of (IN1), (DG1) and (DG2),  one has $ f^2_{12} \ll f_{11}f_{12}$
and $f_{13}^2 \ll f_{11}^2$ and thus the characteristic signature is a
copious production of the final state, $e^\pm\Delta^{\mp\mp}$.
If the low energy decay $\tau\to 3l$ or $\mu\to e\gamma$ is observed,
the value of $f_{11}^2$ is determined by the following comparison with
$f_{11}f_{23}$ and $(ff^\dagger)_{12}$ triggering the decays
$\tau \to \bar{\mu}ee$ and $\mu\to e\gamma$, respectively:
\begin{eqnarray}
f_{11}^2 & =& [2, {4\over R}, 1]\, f_{11}f_{23}
\\
\mbox{or}\quad
f_{11}^2 &= & [{8\sqrt{2} \over 3 r \sin2\theta_3}, {\tt x},
              {4\sqrt{2} \over 3R r \sin2\theta_3}]\, (ff^\dagger)_{12}
\nonumber
\end{eqnarray}
for the cases of (IN1), (DG1) and (DG2), respectively.
Here, ${\tt x}$ cannot be specified as $(ff^\dagger)_{12}$ can be
vanishingly small in the case (DG1).   This shows that $f_{11}^2 \gg 10^{-6}$
and thus the production of $e^\pm \Delta^{\mp\mp}$ can be detected
even for $M_{\Delta^{\pm\pm}} \sim 1$ TeV.
Even in the case that only $\mu\to 3e$ decay is observed, there is some
allowed parameter space for the production of $e^\pm \Delta^{\mp\mp}$
as we have
\begin{equation}
 f_{11}^2 = [{\sqrt{2}\over s_2-{r\over4}\sin2\theta_3},
             {2\sqrt{2}\over R(s_2+{r\over2}\sin2\theta_3)},
             {1\over \sqrt{2}(s_2-{r\over4}\sin2\theta_3)}]\, f_{11}f_{12}
\end{equation}

For the case of (HI),  we have
\begin{eqnarray}
(f_{11}^2,\, f_{13}^2) &=& (t_3^2,\, 2)\, r \sin^22\theta_3 \, f_{22}f_{23}
\nonumber\\
\mbox{or}\quad
(f_{11}^2,\, f_{13}^2) &=& (t_3^2,\, 2)\, \sqrt{r\over2}
\sin2\theta_3 \, (ff^\dagger)_{12}
\end{eqnarray}
when $f_{12}$ is made small to suppress the decay $\mu \to 3e$.
This shows that the decay $\tau\to 3\mu$ and $\mu\to e\gamma$
could be observed together with the collider signals of producing
the events $e\Delta$ and $\tau\Delta$  satisfying the relation
$$ N(e\Delta): N(\tau\Delta) \approx   t_3^2 : 2\,. $$
Let us note that no signal of $l\Delta$ production can be
observed if only the decay $\mu \to 3e$ is observable in the case (HI).

\noindent
$\bullet$ Pair production of $\Delta^{\pm\pm}$:
$\gamma^*,Z^* \to \Delta^{++}\Delta^{--}$.

When the couplings $f_{ij}$ are much smaller than the electroweak gauge
couplings, which is always the case except for (DG1),  pairs of doubly
charged Higgs bosons can be produced
through the gauge interactions exchanging $\gamma$ or $Z$,
if allowed kinematically.  Then, the produced $\Delta^{\pm\pm}$ may decay
mainly to a pair of same-sign charged leptons through the couplings $f$.
In this case, we can measure the relative sizes of
the branching ratios $B(\Delta^{--} \to l_i l_j)$ and thus the
ratios of $f_{ij}$, which enables us to confirm what neutrino mass texture
is realized in nature.  Let us show the expected ratio of
$B(ee): B(\mu\mu) : B(\tau\tau) : B(e\mu):B(e\tau):B(\mu\tau)$
calculated from Eqs.~(6)-(15);
\begin{eqnarray}
\bullet &\mbox{(HI)}~ & 2 r \sin^4\theta_3: {1\over2}:  {1\over2}:
    {1\over2} r \sin^22\theta_3: {1\over2} r \sin^22\theta_3: 1
    \nonumber\\
\bullet &\mbox{(IN1)}~&  1 : {1\over4}:  {1\over4}
        : {1\over16} r^2\sin^22\theta_3:
                {1\over16} r^2\sin^22\theta_3: {1\over2}
    \nonumber\\
\bullet &\mbox{(IN2)}~& \cot^22\theta_3: {1\over4}\cot^22\theta_3:
     {1\over4}\cot^22\theta_3: 1: 1:  {1\over2}\cot^22\theta_3
     \nonumber\\
\bullet &\mbox{(DG1)}~&  1: 1: 1: {1\over16} R^2r^2\sin^22\theta_3:
         {1\over16} R^2r^2\sin^22\theta_3: {1\over8}R^2
\\
\bullet &\mbox{(DG2)}~&  {1\over2}: {1\over32}R^2:  {1\over32}R^2:
 {1\over8}r^2\sin^22\theta_3: {1\over8}r^2\sin^22\theta_3: 1
 \nonumber\\
\bullet&\mbox{(DG3)}~& \cot^22\theta_3: {1\over4}\tan^2\theta_3:
    {1\over4}\tan^2\theta_3: 1: 1: {1\over2} \cot^2\theta_3
    \nonumber\\
\bullet&\mbox{(DG4)}~& \cot^22\theta_3: {1\over4}\cot^2\theta_3:
    {1\over4}\cot^2\theta_3: 1: 1: {1\over2} \tan^2\theta_3
    \nonumber
\end{eqnarray}
In the above expressions, we assumed that $s_2$ is negligible.

   In the linear collider  with $\sqrt{s}=1$ TeV, the pair production
   cross section is
   $\sigma\approx (100-10)$ fb for $M_{\Delta^{\pm\pm}} 
   = (100-450)$ GeV \cite{rizzo}.
   Thus, taking $L=1000$/fb, the number of the produced $\Delta^{\pm\pm}$ will
   be  $N=(10^5-10^4)$.
   In LHC with $L=1000/$fb,  the number of the
   reconstructed pair production  events is expected to be $N=(10^5-10^3)$
   for $M_{\Delta^{\pm\pm}} = (100-450)$ GeV and it becomes down to $N=10$ 
   for $M_{\Delta^{\pm\pm}} =1000$ GeV \cite{lhc}.
Thus, both the linear collider and LHC can produce enough numbers of
$\Delta^{\pm\pm}$ to probe the neutrino mass pattern if $M_{\Delta^{\pm\pm}}
\lesssim 450$ GeV.  In this case, the precise measurement of the branching
ratios can also determine the neutrino oscillation parameters such as
$ r,R$ or $\theta_3$.  It is amusing to note that 
LHC has a good potential to confirm
the triplet Higgs model as the source of neutrino mass matrix
up to the triplet mass around 1 TeV.  For this, the observation of 
the leading decay modes will be enough to 
discriminate the neutrino mass patterns as follows:
\begin{eqnarray}
\bullet &\mbox{(HI)}~ &B(\mu\mu):B(\tau\tau):B(\mu\tau)={1\over2}:{1\over2}:1
 \nonumber\\
\bullet & \mbox{(IN1)}~ &B(ee):B(\mu\mu):B(\tau\tau):B(\mu\tau)=1:{1\over4}
                 :{1\over4}:{1\over2}
          \nonumber\\
\bullet & \mbox{(IN2)}~ &B(e\mu):B(e\tau)= 1: 1
 \nonumber\\
\bullet & \mbox{(DG1)}~ &B(ee):B(\mu\mu):B(\tau\tau)=1:1:1 \\
\bullet & \mbox{(DG2)}~ &B(ee):B(\tau\tau)=1:1
 \nonumber\\
\bullet & \mbox{(DG3)}~ &B(e\mu):B(e\tau):B(\mu\tau)=
     1:1:{1\over2}\cot^2\theta_3
      \nonumber\\
\bullet & \mbox{(DG4)}~ &B(\mu\mu):B(\tau\tau):B(e\mu):B(e\tau)=
     {1\over4} \cot^2\theta_3: {1\over4} \cot^2\theta_3:
     1:1
      \nonumber
\end{eqnarray}
Here we assumed that $\cot2\theta_3$ and $\tan\theta_3$
sit at their lowest allowed values and thus give a sub-leading effect.

\section{A Model: Two-loop generation of  $LL\Delta$ and $\Phi\Phi\Delta$}

An unnatural feature of the Higgs triplet model generating the neutrino
mass is that the model requires another hierarchy of couplings;
the smallness of  $f$ or $\mu$. This would have the same origin as
the hierarchies of the usual quark and lepton Yukawa couplings, which
is one of the difficult problems in particle physics.
In this section, we separate the neutrino sector
from the other and try to explain the smallness of $f$ or $\mu$ through
a radiative mechanism.  In the case of  $f \gg \mu$, a way to get
the small $\mu$ has been explored in Ref.~\cite{babu2} in which
the operator $\Phi\Phi\Delta$ has been obtained at two loop.
A variant of such a scheme can be found to explain the smallness
of both $f$ and $\mu$.  For this, let us introduce the following
new scalar fields and a $Z_3$ discrete symmetry;
\begin{equation}
\begin{array}{cccc}
  X_T & X_Q & X_u & S \cr
(3,3,-{1\over3})_1 & (3,2,{1\over6})_{\alpha^2} & (\bar{3}, 1, 
-{2\over3})_1 & (1,1,0)_\alpha \cr
\end{array}
\end{equation}
where the $SU(3)_c\times SU(2)_L \times U(1)_Y \times Z_3$ charge 
of each field is specified in the second line and $\alpha=e^{2\pi/3}$.  
We assign the $Z_3$ charge $\alpha$  to $L$ and  $\alpha^2$ to  $e^c$ 
and $\Delta$. All the other fields are neutral under $Z_3$.
The allowed couplings are
\begin{equation}
QQ X_T,\quad L d^c X_Q,\quad d^c d^c X_u,\quad X_Q X_Q X_T S^*,\quad \Delta X_T X_u  S.
\end{equation}
Then the operators $LL\Delta S^2$ and $\Phi\Phi\Delta S$ arise from 
the two-loop diagrams as in Figure 1 and thus the small values of $f$ 
and $\mu$ can obtained when $S$ gets a VEV of the order $v_\Phi$.

    \begin{figure} 
    \begin{picture}(200,120)(0,0)
    \DashCArc(100,60)(50,0,180){3} \GBoxc(100,110)(5,5){0}
    \CArc(100,60)(50,180,360) \Vertex(100,10){2}
    \DashLine(100,110)(100,10){3}
    \DashLine(100,60)(130,60){3}  \GBoxc(100,60)(5,5){0}
    \Line(20,60)(50,60)  \Vertex(50,60){2}
    \Line(150,60)(180,60) \Vertex(150,60){2}
    \Text(30,63)[b]{$L$} \Text(170,63)[b]{$L$} \Text(120,63)[b]{$\Delta$}
    \Text(65,100)[r]{$X_Q$} \Text(135,100)[l]{$X_Q$}
    \Text(65,20)[r]{$d^c$} \Text(135,20)[l]{$d^c$}
    \Text(95,85)[r]{$X_T$} \Text(95,35)[r]{$X_u$}
    \end{picture}
    \begin{picture}(200,120)(0,0)
    \CArc(100,60)(50,0,180) \Vertex(100,110){2}
    \CArc(100,60)(50,180,360) \Vertex(100,10){2}
    \DashLine(100,110)(100,10){3}
    \DashLine(100,60)(130,60){3}  \GBoxc(100,60)(5,5){0}
    \DashLine(20,60)(50,60){3}  \Vertex(50,60){2}
    \DashLine(150,60)(180,60){3} \Vertex(150,60){2}
    \Text(30,63)[b]{$\Phi$} \Text(170,63)[b]{$\Phi$} \Text(120,63)[b]{$\Delta$}
    \Text(65,100)[r]{$Q$} \Text(135,100)[l]{$Q$}
    \Text(65,20)[r]{$d^c$} \Text(135,20)[l]{$d^c$}
    \Text(95,85)[r]{$X_T$} \Text(95,35)[r]{$X_u$}
    \end{picture}  \\ 
    \caption{Two loop diagrams generating the operators $LL\Delta$ and 
    $\Phi\Phi\Delta$.  Black squares represent vertices with 
    $\langle S \rangle$.} 
    \end{figure}
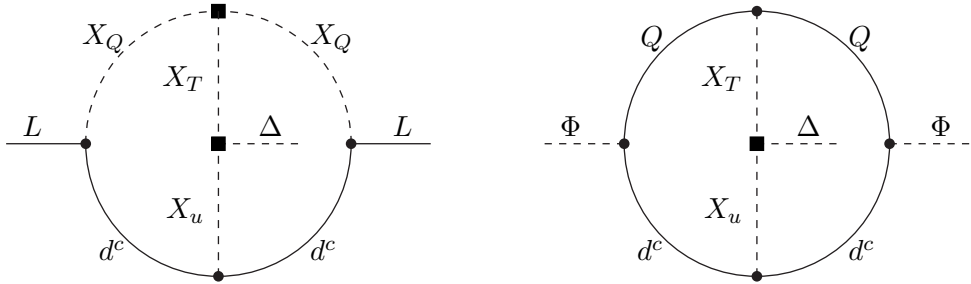

\section{conclusion}

We have investigated the testability of the low-energy Higgs triplet 
model and the resulting neutrino masses and mixing 
in the future collider experiments.  The bileptonic couplings 
$f_{ij}$ can be large enough to yield observable
lepton flavor violating decays of a charged lepton such as 
$\mu\to 3e, \mu\to e\gamma$ or $\tau \to 3l$ depending  
on the neutrino mass patterns.  For this to happen, 
the coupling $f_{12}$ needs to be vanishingly small in order 
to satisfy the current bound on the $\mu\to 3e$ decay.
Another effect of the bileptonic couplings is the production of  a
doubly charged Higgs boson accompanied by a charged lepton $l_i$
in the $e^+ e^-$ collider. In this case,
we have identified the characteristic flavor structure of the final 
state, $l_i^\mp \Delta^{\pm\pm}$, for each neutrino mass pattern.
 We have shown that  copious production of the doubly charged 
 Higgs boson pairs through the gauge interactions 
 in the linear collider and LHC provides a promising 
 way to test not only the triplet Higgs model but also the resulting
 neutrino mass matrix even when $f$ is very small.
 In LHC, in particular, we expect sufficient production of 
 the doubly charged Higgs bosons up to the mass $\sim 1$ TeV which
 will enables us to determine the neutrino mass pattern only 
 by observing the leading decay channels.
A problem in the low-energy triplet Higgs model is how to understand 
the smallness of the couplings $f$ and $\mu$.  We have also worked out 
a radiative mechanism as one of possible solutions.

\medskip
\noindent
{\bf Acknowledgment}:
EJC was supported by the Korea Research Foundation Grant,
KRF-2002-015-CP0060.


\begin{thebibliography}{99}

\def\plb#1#2#3{Phys.\ Lett.\       {\bf B#1}  (#2) #3}
\def\npb#1#2#3{Nucl.\ Phys.\       {\bf B#1}  (#2) #3}
\def\prd#1#2#3{Phys.\ Rev.\        {\bf D#1}  (#2) #3}
\def\prl#1#2#3{Phys.\ Rev.\ Lett.\ {\bf #1}   (#2) #3}
\def\mpl#1#2#3{Mod.\ Phys.\ Lett.\ {\bf A#1}  (#2) #3}
\def\rep#1#2#3{Phys.\ Rep.\        {\bf #1}   (#2) #3}
\def\sci#1#2#3{Science             {\bf #1}   (#2) #3}
\def\astro#1#2#3{Astrophys.\ J.\   {\bf #1}   (#2) #3}
\def\epj#1#2#3{Eur.\ Phys.\ J.\   {\bf C#1}   (#2) #3}
\def\jhep#1#2#3{JHEP              {\bf #1}   (#2) #3}
\def\ptp#1#2#3{Prog.\ Theor.\ Phys.\ {\bf #1}  (#2) #3}

\bibitem{SK}
 Super-K Collaboration, Y. Fukuda {\it et. al.}, \prl{81}{1998}{1562}.
\bibitem{SNO}
 SNO Collaboration, Q.R. Ahmad {\it et. al.}, \prl{89}{2002}{011301}.
\bibitem{CHZ}
  CHOOZ collaboration, M. Apollonio {\it eta. al.}, \plb{420}{1998}{397}.
\bibitem{KamL}
 KamLAND Collaboration, K. Eguchi {\it et. al.}, \prl{90}{2003}{021802}.
\bibitem{rpv}
  B. Mukhopadhyaya, S. Roy and F. Vissani, \plb{443}{1998}{191};
  E.J. Chun and J.S. Lee, \prd{60}{1999}{075006}; S.Y. Choi {\it et. al.},
  \prd{60}{1999}{075002}; W. Porod {\it et. al.}, \prd{63}{2001}{115004}.
\bibitem{babu1}
 A. Zee, \plb{93}{1980}{389}; 
 K.S. Babu and C. Macesanu, hep-ph/0212058.
\bibitem{gunion}
 J.F. Gunion, J. Grifols, A. Mendez, B. Kayser and F. Olness,
 \prd{40}{1989}{1546};
 R. Vega and D. Dicus, \npb{329}{1990}{533};
 J.F. Gunion, R. Vega and J. Wudka, \prd{42}{1990}{1673};
\bibitem{xyz}
 R. Godbole, B. Mukhopadhyaya and M. Nowakowski, \plb{352}{1995}{388};
 K. Cheung, R. Phillips and A. Pilaftsis, \prd{51}{1995}{4731};
 K. Huitu, J. Maalampi, A. Pietila and M. Raidal, \npb{487}{1997}{27}.
\bibitem{rizzo}
 T.G. Rizzo, \prd{45}{1992}{42};
 N. Lepore, B. Thorndyke, H. Nadeau and D. London, \prd{50}{1994}{2031}.
\bibitem{lhc}
 J.F. Gunion, C. Loomis and K.T. Pitts, hep-ph/9610237;
 B. Dion {\it et. al.}, \prd{59}{1999}{075006};
 A. Datta and A. Raychaudhuri, \prd{62}{2000}{055002}.
\bibitem{ma1}
 E. Ma, M. Raidal and U. Sarkar, \prl{85}{2000}{3769}; \npb{615}{2001}{313}.
\bibitem{olds}
 For a review, see,
 F. Cuypers and S. Davidson, \epj{2}{1998}{528}.
\bibitem{wmap}
 D.N. Spergel {\it et.al.}, astro-ph/0302209.
\bibitem{pdg}
 K. Hagiwara {\it et al.} (PDG), Phys. Rev. D66, 010001 (2002) 
\bibitem{yusa}
 Y. Yusa, {\it et.al.}, Belle Collaboration, hep-ph/0211017.
\bibitem{fut1}
  Y. Kuno and Y. Okada,  Rev.~Mod.~Phys.~{\bf 73} (2001) 151.
\bibitem{babu2}
 K.S. Babu and C.N. Leung, \npb{619}{2001}{667}.

\end{thebibliography}
\end{document}